\documentclass[aps,preprint]{revtex4}%
\usepackage{amssymb}
\usepackage{amsfonts}
\usepackage{amsmath}
\usepackage{graphicx}%
\providecommand{\U}[1]{\protect\rule{.1in}{.1in}}

\begin{document}
\preprint{ }
\title{Time and Consciousness in a Quantum World.}
\author{Augusto Cesar Lobo}
\affiliation{Department of Physics-Federal University of Ouro Preto-Brazil}
\email{lobo@iceb.ufop.br}
\keywords{Foundations of quantum mechanics, time and consciousness}
\begin{abstract}
We address the relation between two apparently distinct problems: The quest
for a deeper understanding of the nature of consciousness and the search for
time and space as emergent structures in the quantum mechanical world. We also
advance a toy-model proposal of emergence of time from a timeless unus mundus
quantum-like space by using Aharonov's two state formalism of quantum
mechanics. We further speculate on these issues within a quantum cognitive
perspective with particular interest in two recent papers on this emerging
field of science. One (Aerts et al) entails (as we argue) a panpsychist
top-down approach to the problem of consciousness. The second paper (Blutner
et al) proposes a quantum cognitive model for Jung's psychological type
structure. We discuss these concepts and their relation with our main thesis,
that \textit{time is a measure of individuality}. One of our central
motivations is to provide arguments that allows the mainstream physicist to
take seriously a panpsychist worldview, a position that has been openly
forwarded by many modern philosophers.

\end{abstract}
\maketitle
\tableofcontents

\section{Time in Quantum Theory as a Measure of Individuality}

According to Kant, space and time are \textit{a priori} non-empirical
representations and as such, they underlie all human mental constructions in
order to organize and apprehend the sensorial data of physical reality
\cite{kant1998critique}. One could make the case that Einstein's relativity
theory undermined this notion. We know since the early twentieth century that
the inner workings of space and time can differ radically from the structure
that our naive intuition leads us to. The Minkowski structure of spacetime was
far from obvious and was discovered by studying carefully the discrepancy
between the structure of electromagnetism (spawned by a large body of
empirical evidence) and the Newtonian concept of space and time
\cite{einstein1989collected}, \cite{minkowski1952space}. One could well say
that Maxwell's equations already contains in itself the Lorentz spacetime
structure and that Einstein and Minkowski were the first to fully understand
this fact.

In the third decade of the twentieth century, the theory of a non-relativistic
quantum theory (NRQT) of particles was developed by names as Heisenberg,
Schr\"{o}dinger, Born, Bohr, Jordan and Dirac \cite{weinberg1996quantum}. This
theory is \textit{non-relativistic} in a serious way. Time and space are
treated in radically distinct ways: Time is a one-dimensional
\textit{parameter} and space is an \textit{observable}, an Hermitian operator
defined on an abstract infinite-dimensional quantum space of state-vectors. Of
course, it was immediately recognized by the founders of quantum physics that
it was imperative to extend the theory in order to\ make it compatible with
relativity. Yet, the route to this approach was not so clear in the beginning.
For instance, when Bohr was told that Dirac was working on a relativistic
equation for the electron, he was surprised because he thought that Klein and
Gordon had already achieved this \cite{jammer_philosophy_1974}.

After a while, it was recognized that no single particle equation would do the
trick. The concept of a relativistic quantum field theory (RQFT) was
introduced in order to describe elementary particles and their interactions.
Particles were interpreted as elementary excitations of these quantum fields.
The position operators of NRQT were "downgraded" to a simple parameter along
with time in Minkowski's spacetime. And there were good reasons for this.
First, Dirac observed that the \textit{opposite} move (upgrading time as an
observable) did not make much sense because then time and energy would both
have semibounded by below spectrum like position and momentum observables. But
this clearly conflicts with the stability of fermionic quantum systems as the
hydrogen atom for instance. Yet, it was precisely this lack of stability of
classical models for atomic systems that was actually one of the major driving
forces behind the development of quantum mechanics in the first place. A
second physical argument, consistent with the first can be made. The concept
of a particle state with a well defined position does \textit{not} make sense
because of the intrinsic lack of an operational implementation of the
measurement of its position with arbitrary precision. (This is only
\textit{one instance} of an "operational-positivistic" argument that pervades
the entire history of quantum physics). For example, consider an electron
interacting with a classical electromagnetic (EM) field. In order to
\textit{confine} the electron inside a wave-packet with a width $\Delta x$
smaller than its Compton length $1/m_{e}$ (in natural units) one needs to set
an EM field so strong that it would unavoidably create positrons and electron
pairs implying a many-body formalism with a varying number of particles
\cite{landau1971course}. This provides a powerful reason for a quantum field
description of particle interactions. Even nowadays, our most refined theories
of physical reality accepted by mainstream physics is based on this
construction. The standard model of elementary particles and even most
attempts to go beyond it (supersymmetry and string theories as primary
examples) are one way or another ultimately defined within this paradigm. Yet,
there is something \textit{odd} about this model. First, this move of the role
played by the position of quantum particles seems strange. Indeed, compare it
to how the non-relativistic limit of relativistic classical mechanics
(classical as non-quantum) is understood. In the latter, the passage from the
relativistic regime to the Newtonian one is straightforward. It can be derived
easily for relative speeds that are \textit{small} compared to the speed of
light. This "structural transformation" of the position operator to a simple
parameter for relativistic quantum field theory is not sufficiently understood
and many definitions of a relativistic position operator have been
sporadically proposed in the literature \cite{wang2003time}. A second and much
more serious "positivistic flavoured argument" can been made. Consider a full
RQFT explanation of particle interactions. For example, suppose (for the sake
of simplicity) that the world is comprised only by electrons, positrons and
photons. We have an electronic field and the "photonic field" which mediate
the interaction between the electronic leptons. Given a state of the
electronic/EM field all one can compute and "ask experimentally" are things
like "For \textit{this} particular event of spacetime, what is the probability
of finding a certain number of electrons, positrons or photons with such and
such momenta, spin, helicity, etc.?" The typical scattering experiment is one
where there is a \textit{global} field state characterized by a given number
of \textit{incoming} particles far away with well defined momenta, spin, etc.
and \textit{outgoing} particles (long after the interaction) again with a well
defined number of particles with given states. One then computes and measures
the transition probability. What is curious about this full-blown quantum
field description is that since the particles are \textit{only excitations} of
quantum fields, there is \textit{no} room for any\textit{ individuality} of a
single particle. This is a well-known fact, but it seems that the full depth
of the fundamental consequences of this issue has not been pursued far enough.
Take, for instance, a simple non-relativistic quantum system as the hydrogen
atom. In this case, one has an electron quantum mechanically bound to a proton
by its classical electric field. The electron can be in \textit{this} or
\textit{that} energy state with \textit{this} or \textit{that} spin. But we
think of the electron as being the \textit{same} particle that happens to be
in different possible states over time. From the point of RQFT, this is a
"linguistic liberty", so to speak. What \textit{really} happens (according to
RQFT) is that the (classical) EM field in this case is not strong enough to
create more particles/excitations with appreciable probability, thus, the
individuality of this particular field excitation is \textit{emergent}. This
poses the following question: how do we measure time operationally? The
obvious answer is: with a \textit{clock}. But a clock is nothing else but a
periodic phenomenon of some sort. For instance, the rotation of the Earth
around its axis or its translational orbit around the sun were once considered
excellent clocks. Today we apply atomic clocks to GPS technology. A necessary
condition in order to an arbitrary system be considered a clock is that it
must maintain its \textit{individuality}. If an object now is supposed to be a
clock, one must be able to recognize this system as the \textit{same} object
in a further moment. How is it then that we have a full-blown RQFT with a
previously defined spacetime? Einstein changed our understanding of spacetime
when he thought carefully about how one can operationally define the
synchronization of clocks in a way compatible with electromagnetism. It seems
that what quantum mechanics is telling us about the nature of spacetime is
that its structure should be \textit{emergent} in some strong ontological
sense that still remains to be unravelled. The structure of RQFT is a
"superstructure" in this sense. Time and space are artificially included "by
hand" in the theory. For this reason it is natural to claim that, in some
sense to be made more precise in the future, one can state that "\textit{Time
is a Measure of Individuality}". We address a few suggestions of how this may
be accomplished in the following sections. In the last section, we conclude
with some closing remarks. We visit now a concept that is possibly even more
controversial than that of the nature of time or spacetime.

\section{ Time and Consciousness: The true "astonishing hypothesis"}

Francis Crick, one of the 1962 Medicine nobel-prize winners for co-discovering
the double-helix structure of DNA (together with James Watson and Maurice
Wilkins) published a book in 1994 called "\textit{The Astonishing Hypothesis:
The Scientific Search for the Soul}" \cite{crick1994astonishing}. In his book,
he advocates a physicalist and reductionist explanation of the mind and other
attributes of consciousness. From the perspective of the History of Science,
Newtonian Physics was followed by an increasingly number of new developments
over the past three or four centuries together with amazing advances of the
hard and biological sciences. It would seem that one could conclude today that
the reductionist point of view is completely vindicated. From a mainstream
scientific standard, all attempts of a vitalist philosophy have been purged
from science and as so, the reductionist hypothesis (sometimes called the
\textit{bottom-up} approach) for explaining consciousness actually looks
anything but astonishing.

It fact, it is the opposite approach, the hypothesis that the concept of
consciousness is somehow a primitive ontological feature of reality is
nowadays what deserves to be called \textit{astonishing}. If one looks at the
history of science, though, a much more intertwined complex intellectual
zigzag between both metaphysical positions becomes evident. Some of the most
important founders of the age of reason were not so convinced themselves of
this extreme form of "naive physicalism". Descartes advanced his well-known
dualistic approach, where mind can affect matter, but not the contrary
\cite{descartes1985description}. His point of view is now considered almost
universally unacceptable given our present knowledge of science. Neither did
Newton believe in a completely mechanical universe that obeys deterministic
laws (that were essentially discovered by himself.) As a matter of fact, he
somehow was able to maintain a metaphysical view with enough room to
accommodate a God that could once in a while "intervene" in the workings of
the great machine \cite{newton1979opticks}. Leibniz argued in his
"\textit{Monadology}" that there is an inescapable "explanatory gap" between
the purely mechanistic explanation of the world and the true functioning of
the "mind" \cite{leibniz1989monadology}. Yet, many of Newton's later followers
become much more "Newtonian" than himself. Take Laplace's famous words
\cite{laplace2012pierre}:

"\textit{We may regard the present state of the universe as the effect of its
past and the cause of its future. An intellect which at a certain moment would
know all forces that set nature in motion, and all positions of all items of
which nature is composed, if this intellect were also vast enough to submit
these data to analysis, it would embrace in a single formula the movements of
the greatest bodies of the universe and those of the tiniest atom; for such an
intellect nothing would be uncertain and the future just like the past would
be present before its eyes}."

Needless to reaffirm, this increasing hard-core belief in \textit{mechanism}
as an ultimate description of reality was supported by an extraordinary number
of successful applications of physics and mathematics to all kinds of natural
phenomena. Yet, a reaction against this prevailing attitude from the
intellectual world did not have to wait much: already in the late eighteenth
century, a number of different schools of thought (much later collectively
coined as the \textit{Counter-Enlightment} movement) advocated a more
anti-rationalist worldview where vitalist and organic ideas were welcomed as a
cure for the excessive cold and mechanistic metaphysics of the times. These
ideas were closely related to German Romanticism and many thinkers and artists
(Goethe for instance) defended many anti-enlightment positions
\cite{berlin2013counter}. We dwell into these historical upturns only to point
out that the resurgence of panpsychism and monist related philosophies in the
last few decades are far from being something new in the history of science.
In the twentieth century, many founders of quantum mechanics also dived into
some related concepts. Bohr, Schr\"{o}dinger, von-Neumann, Wigner and Pauli
all considered the idea of consciousness (or significantly close ideas)
playing a fundamental role in quantum mechanics. Bohr introduced the
philosophical concept of \textit{complementarity }in order to deal with
apparently logical contradictions found in quantum descriptions as the famous
wave-particle duality. He believed that this concept could be extended to many
other fields far from quantum physics as the complementarity between science
and religion, for example. Bohr was also convinced that the so called
\textit{measurement problem} could be solved by imposing that a state vector
of a quantum system collapses when it interacts with a classical measuring
system. Thus, in his view, classical reality should have an ontological status
equal to quantum reality. But the \textit{circular logic} implied by the fact
that classical reality should also obviously be some kind of\textit{ classical
limit} of the quantum description of the world did not seem to bother him at
all. Bohr was also influenced by Taoist philosophy. Indeed, when Frederick IX
conferred him the Order of the Elephant, he designed his own coat of arms
which featured the \textit{yin/yang} symbol together with the Latin motto
\textit{contraria sunt complementa}, "opposites are complementary"
\cite{paiz1992niels}. Schr\"{o}dinger was pretty much influenced by the
philosophy of Schopenhauer and like him also had strong interests in Eastern
philosophies. In his book \textit{What Is Life?} (curiously the book that
decisively influenced Watson and Crick in their pursuit of the structure of
DNA) he speculates about the possibility that individual consciousness could
be a manifestation of some kind of universal consciousness
\cite{magee1997philosophy}, \cite{schrodinger1992life}. It is also well-known
that both von-Neumann and Wigner suggested the introduction of a
\textit{conscious observer} in order to solve the measurement problem in
quantum mechanics \cite{wigner1995remarks}. Yet, it is fare to say that it was
Pauli and his friend Jung that pushed these ideas to an unprecedented level
\cite{laurikainen2012beyond}.

Jung was a Swiss psychiatrist who founded the discipline of analytical
psychology. One of its main concepts is that of \textit{individuation.} a
process of integrating the conscious realm with the unconscious one in order
to provide a healthy human development of the psyche. He introduced a number
of concepts like \textit{psychological types}: the extrovert and introvert
\textit{attitudes} together with the \textit{psychological} \textit{functions}
of Sensation, Intuition, Thinking and Feeling. He also developed the concept
of archetypes, collective unconscious, the ego and the shadow, the animus and
the anima and \textit{synchronicity}. This last concept involved the idea of
non-causal relations between events with psychological significance. This
greatly influenced Pauli (that was his patient) and they become collaborators
and friends. They developed together the notion of the \textit{unus mundus} in
connection to non-local and non-causal synchronistic phenomena. The
\textit{unus mundus} can be thought as a deep "undifferentiated sea" of
unconscious possibilities where common elementary archetypes of the human race
resides and that occasionally surfaces through the particular individuation of
human beings.

Jung developed these conceptual constructs after many years both of clinical
observations of his patients and personal introspection. Both thinkers
believed that the \textit{objective} world and the \textit{subjective} world
were equally \textit{real} and important and that a proper understanding of
the relation between these distinct ontological realms was a matter of great
urgency. They were convinced that physics (and science in general) would
evolve towards this path in the future.

These indeed are extraordinary and truly astonishing claims. After all, the
beginning of modern science was shaped during the Renaissance and the
Enlightenment when a clear cut between objective and subjective reality was
devised. How could it even be possible to consider going back to the
superstitious, religious and "magical thinking" of medieval times? Today, most
thinkers that consider panpsychism seriously are considered as
anti-rationalists by mainstream physicalists \cite{dawkins2000unweaving}.
Probably a superficial "new-agism" together with some exaggerated post-modern
relativism did not help much in this matter \cite{sokal1998intellectual},
\cite{arntz2007bleep}. The work of Philosophers and Cognitive Scientists
varying from Gregory Bateson to David Chalmers, Liane Gabora and Ignazio
Licata are, unfortunately, not sufficiently appreciated, in general, by the
current hard materialistic paradigm \cite{bateson1972steps},
\cite{chalmers1996conscious}, \cite{gabora2002amplifying},
\cite{licata2007physics}.

We distinguish \textit{four }main distinct philosophical positions on this
issue. To make this point clear, let us quote a famous line of Wittgenstein
from his "\textit{Tractatus Logico-Philosophicus}": "\textit{Whereof one
cannot speak, thereof must one be silent}." \cite{wittgenstein_tractatus_1999}
The \textit{first view} is that of the pure hard-core materialistic and
physicalist thinker. He would interpret Wittgenstein's words as meaning that
all there is about reality are precisely \textit{only} those things that can
be said about it. This position represents such a naive view that it almost
excuses us from further commentaries. It is enough to say that this represents
precisely the opposite of what Wittgenstein was trying to communicate
\cite{janikwittgenstein}. This impoverished philosophical stand is commonly
found to be behind a certain kind of revived old-fashioned discourse coined
pejoratively as "scientism". Yet, many contemporary mainstream scientists go
along with this metaphysical position . The \textit{second view} is probably
much more akin to Wittgenstein's original vision. This is the view of the
mainstream scientist that does \textit{not} negate the existence of things in
the world that are \textit{beyond} science, but believes that since these
matters belong (by definition) \textit{out} of the domain of scientific
discourse, they should be dealt exclusively by non-scientific disciplines like
ethics, religion, etc. This is a respectable and pragmatic position and there
are good reasons for taking this metaphysical position seriously. A
\textit{third view} envisages a philosophy of monism in order to describe
reality where the nature of consciousness and that of material reality are one
and the same. This philosophy of panpsychism can be seen as a
\textit{top/bottom} approach to the mind/body problem where consciousness is a
primary ontological feature of the world and where any element of material
reality has at least some degree of consciousness \cite{wendt2015quantum}.
Yet, there is a \textit{fourth view} that can be considered as being somewhat
\textit{between} views number two and number three. This stems from the fact
that one may formulate the following question (about view number two): if
there is a clear division between those things that we can speak about and
those that we cannot, how can we talk about the division itself? How can one
recognize it? One possible answer is that we talk about the "unspeakable"
\textit{indirectly} through metaphors, art, religion and cultural expressions
in general. But, the cultural forms and language that we use in order to
express science (and in particular, physics) also \textit{evolves in time}. As
an example, take Faraday's concept of \textit{force fields} that he introduced
in the nineteenth century. This represented a major step away from the local
particle interaction model of Newtonian mechanics. It was difficult for the
physicists to understand this new form of thinking. It was Hertz that finally
accepted the fact that Maxwell equations \textit{were} the EM theory, paving
the way for Einstein. All the other main physicists of the day (including
Maxwell himself) tried to construct mechanical models for a luminiferous
ether. Today any child is familiar with the idea of a "force field". It is
commonly depicted in many contemporary cartoons, TV shows and science-fiction
movies. Who knows what people will think about \textit{wave-functions} and
\textit{quantum entanglement} in a couple of hundred years from now? Richard
Feynman in his famous \textit{Lectures on Physics} that he delivered back in
the sixties to undergraduate students at Caltech has a section named
"\textit{Scientific imagination}". Some excerpts from the original text are
quoted below \cite{feynman1979feynman}:

"...\textit{I have no picture of this electromagnetic field that is in any
sense accurate. ...When I start describing the magnetic field moving through
space, I speak of the }$\vec{E}$\textit{ and }$\vec{B}$\textit{ fields and
wave my arms and you may imagine that I can see them. I'll tell you what I
see. I see some kind of vague shadowy, wiggling lines---here and there is an
}$\vec{E}$\textit{ and }$\vec{B}$\textit{ written on them somehow, and perhaps
some of the lines have arrows on them---an arrow here or there which
disappears when I look too closely at it. When I talk about the fields
swishing through space, I have a terrible confusion between the symbols I use
to describe the objects and the objects themselves. ...We use a lot of tools,
though. We use mathematical equations and rules, and make a lot of pictures.
What I realize now is that when I talk about the electromagnetic field in
space, I see some kind of a superposition of all of the diagrams which I've
ever seen drawn about them...}

\textit{...Perhaps the only hope, you say, is to take a mathematical view...We
are unfortunately limited to abstractions, to using instruments to detect the
field, to using mathematical symbols to describe the field, etc. But
nevertheless, in some sense the fields are real, because after we are all
finished fiddling around with mathematical equations---with or without making
pictures and drawings or trying to visualize the thing---we can still make the
instruments detect the signals from Mariner II and find out about galaxies a
billion miles away, and so on."}

Feynman gives us here a vivid and pedagogical example of how striking the
influence of historical and cultural contexts can exert upon scientific discourse.

It is our firm belief that there in \textit{no such thing} as a completely
\textit{neutral} philosophical thinker. Everyone carries their own prejudices
and bias. At this point, the reader probably has guessed that the author has
sympathies towards \textit{views} number three and four. Yet, it is absolutely
imperative to recognize that there is presently \textit{no} hard scientific
facts capable of distinguishing any of these metaphysical positions. In fact,
for view \textit{number two} thinkers, it is even impossible to distinguish
them in principle. The \textit{view} number one physicalist (clearly also a
kind of an extreme monistic and materialistic thinker) believes that the
"explanatory gap" can be closed when the workings of the human brain become
sufficiently understood someday in the future.

We propose some concrete pathways where one can look for a intellectual
construct that can accommodate \textit{views} three or four. Firstly it is
important to acknowledge that there is a large and ancient historical body of
evidence of knowledge from the "subjective realm". This may seem a
contradiction in terms because we are used to think that only objective facts
are capable of being communicated and that deserves being recognized as
science. Take physics and mathematics as a typical example. The laws of Newton
are the same in Europe or Asia and relativity theory is the same in the
Northern or Southern hemispheres. The famous number theorists Hardy and
Ramanujan could work perfectly together on mathematics in the early twentieth
century, but the first was a typical European rationalist and the second was a
religious Hindu that believed that his uncanny mathematical skills were
delivered to him by his family goddess Mahalakshmi \cite{kanigel2015man}. Yet,
"softer" sciences (as sociology and psychology, for instance) are much more
"cultural dependent" than it is for "hard sciences". Still, their disciplines
are commonly considered as part of the scientific enterprise in general.

We put forward the idea of looking more carefully to what is called
"\textit{perennial philosophy}", a term introduced by Agostino Steuco and
later by Leibniz. Modernly, this concept was recovered by names as Jung,
Joseph Campbell and Aldoux Huxley \cite{huxley2014perennial},
\cite{campbell1990transformations}.

Perennial philosophy\textit{ }can be thought as a tradition which states that
the psychological structure of \textit{consciousness} (different
\textit{states of consciousness} and in particular, \textit{mystical
experiences)} stem from a \textit{universal} concept. Yet, these ideas take
different configurations based on distinct historical and cultural contexts,
thus shaping a variety of schools of thought and organized religions in all
places and times of mankind history. We assume that the concept of
consciousness and states of consciousness (as a working hypothesis) are
\textit{real} in the sense that they have equal ontological status as
mainstream scientific concepts like time, space, energy, matter, etc. The
evidence that allows us to pursue this approach results from a vast and long
time literature based on \textit{introspection} and exploration of subjective
experiences both in Eastern and Western philosophies. Yet, there is only on
true test that can prove if this or any other scientific working hypothesis is
worthwhile. This is the ultimate pragmatic test where the fruitfulness of an
idea is measured by the number of facts it can explain and practical
usefulness it can provide. It should also be consistent with the body of
science that we already know. There is one particular characteristic of these
altered states of consciousness that we are specially interested in. This is
the phenomenon that is sometimes called "\textit{mindfulness}" obtained
through meditation and mind-altering substances. There is a long tradition of
statements that converge to a few ideas about these states: a feeling of a
"loss of ego" and an altered sense of the flux of time. There seems to be a
\textit{continuous spectrum} of states between our \textit{ordinary} state of
consciousness related to a usual sense of psychological time and a complete
loss of ego together with a feeling of "timelessness" for extreme states of
consciousness. For example, a somewhat "lighter" change of the ordinary state
of consciousness is the so-called "day-dreaming". This is a common experience
that usually occurs spontaneously followed by a sense of a lack of passage of
time and immediate memory \cite{weil1986natural}. The chain of thought
processes that happens continuously for all of us when we are in our ordinary
state of consciousness basically "defines" our sense of an "ego" and even a
mild interruption of this process changes our sense of individuality and our
sense of a passage or flux of psychological time. The striking parallel with
our analysis of \textit{physical time} in the previous section is obvious.
Hence, we may rephrase that somehow "\textit{time is a measure of
individuality}". In the next section we deliver a more concrete face to this
idea by studying a quantum mechanical "toy model" that may help shed some
light on this issue.

\section{The Emergence of Time}

\subsection{ The Partial Trace and the Emergence of Temperature}

Feynman discusses in his text-book on Statistical Mechanics
\cite{feynman1998statistical} a concept that is known as the \textit{improper
}density matrix. Suppose one is interested in studying a certain physical
quantum system defined by a finite dimensional space of state-vectors $W_{S}$.
(We consider only finite dimensional spaces in order to avoid inessential
analytical technical details). Feynman then introduces the "rest of the
universe" described by $W_{R}$ so that the "whole universe" is described by
the tensor product $W=W_{S}\otimes W_{R}$. Suppose further that an observer
has physical access only to system $W_{S}$, then given a pure state-vector
$\left\vert \Psi\right\rangle \in W$ of the "universe", one can define a
unique operator $\hat{\rho}_{\left\vert \Psi\right\rangle }$ that acts upon
$W_{S}$ defined by the following equation:
\[
\left\langle \Psi\right\vert \hat{O}\otimes\hat{I}\left\vert \Psi\right\rangle
=tr\left(  \hat{\rho}_{\left\vert \Psi\right\rangle }\hat{O}\right)
\]
for every observable $\hat{O}$ of $W_{S}$. This operational definition singles
out a map between the \textit{space of rays} defined by $W$ and the space of
\textit{positive unit trace} operators in $W_{S}$. One recognizes this map as
the well-known \textit{partial trace} operation
\[
\left\vert \Psi\right\rangle \left\langle \Psi\right\vert \longrightarrow
\hat{\rho}_{\left\vert \Psi\right\rangle }=tr_{R}\left(  \left\vert
\Psi\right\rangle \left\langle \Psi\right\vert \right)
\]
commonly used in modern quantum information theory. Feynman, of course, was
not meaning anything "cosmological" with his choice of words. He meant simply
that given a particular physical system that we are interested in, say a gas
confined by a piston in a cylinder, the "rest of the universe" could be any
amount of the environment relevant to the physics of the system. (A heat bath,
for instance). In order to characterize the system to be in thermal
equilibrium, it is natural to suppose that the density matrix must commute
with the Hamiltonian $\hat{H}$ of the system, so that the density matrix is a
function of the Hamiltonian $\hat{\rho}=\hat{\rho}(\hat{H})$. We further
assume that the entropy $S=-tr(\hat{\rho}\ln\hat{\rho})$ is maximized\textit{
}together with\textit{ }the further constraint defined by the internal energy
$E=tr\left(  \hat{\rho}\hat{H}\right)  $. This leads us to the well-known
Boltzmann-like density operator
\[
\hat{\rho}=\frac{1}{Z}e^{-\beta\hat{H}}\qquad\text{with\qquad}Z=tr(e^{-\beta
\hat{H}})
\]
where $\beta$ is thermodynamically identified with the \textit{inverse
temperature} (in natural units). The Hermitian operator $e^{-\beta\hat{H}}$
looks similar to the unitary time evolution operator $\hat{U}(t)=e^{-it\hat
{H}}$ if one makes the identification $\beta\rightarrow it$. This formal
analogy allows thermal averages of systems in equilibrium to be computed from
RQFT with imaginary time. If a solution to the latter is \textit{analytic} in
time, a thermodynamic solution can be obtained by \textit{Wick rotation}. Thus
it is a mainly a mathematical technique for calculating thermodynamic
partition functions. Speculations that there may be some deeper foundational
reason for these results have been made in the literature but there is not yet
any hard evidence for these claims \cite{zee2010quantum}. Yet, we may
speculate about the possibility of "deriving a emergent notion of time" in an
analogous manner. At a first glance, it seems difficult to find a way to
somehow derive an \textit{unitary} time evolution operator instead of the
Hermitian Boltzmann operator. In the following subsection we suggest how to
circumvent this problem.

\subsection{ Aharonov's Two-State Formalism}

The Two-Time formalism for Quantum Mechanics developed by Aharonov, Bergman
and Lebowtiz (ABL) in 1964 \cite{aharonov1964} was initially proposed in order
to remove the apparent time asymmetry from the usual formulation of Quantum
Physics due to the projection postulate. This formulation advocates that in
order to provide a complete information for a given quantum state of a system
one needs to know not only the previous pre-selected state of the system
obtained by a strong measurement but also a post-selected state given also by
a strong measurement. This is a time-symmetric refinement of the ensemble
given only by a given pre-selected state. This concept led in the eighties to
the discovery of a new \textit{element of reality}, the so-called \textit{Weak
Value }(WV) of an observable for a particular two-time state
\cite{aharonov2008quantum}. Given non-orthogonal pre and post selected states
$\left\vert \alpha\right\rangle $ and $\left\vert \beta\right\rangle $ (the
two-time state is usually represented by the tensor product $\left\langle
\beta\right\vert \otimes\left\vert \alpha\right\rangle $) then the WV
$\left\langle \hat{O}\right\rangle _{w}$ of an observable $\hat{O}$ is the
complex number
\[
\left\langle \hat{O}\right\rangle _{w}=\frac{\left\langle \beta\right\vert
\hat{O}\left\vert \alpha\right\rangle }{\left\langle \beta\right\vert
\alpha\rangle}%
\]

It can be shown that the WV can be effectively measured by conducting a great
number of experiments with the same chosen two-time boundary conditions. This
can be achieved through an infinitesimally small coupling of the system of
interest with a measuring system within a von-Neumann pre-measurement setup.

This concept is indeed time-symmetric and the temporal inversion operation can
be implemented by swapping the pre and post-selected states together with the
complex conjugation of the WV. Consider now the following operationally
definition: take two distinct \textit{maximally entangled} states $\left\vert
\alpha\right\rangle $, $\left\vert \beta\right\rangle $ of the "whole
"universe" $W=W_{s}\otimes W_{R}$ analogously as discussed in the previous
subsection. In this case both $\hat{\rho}_{\left\vert \alpha\right\rangle }$
and $\hat{\rho}_{\left\vert \beta\right\rangle }$ have \textit{maximum}
von-Neumann entropy $S=-tr\left(  \hat{\rho}_{\left\vert \alpha\right\rangle
}\ln\hat{\rho}_{\left\vert \alpha\right\rangle }\right)  =-tr\left(  \hat
{\rho}_{\left\vert \beta\right\rangle }\ln\hat{\rho}_{\left\vert
\beta\right\rangle }\right)  =\ln N$, where $N=\min(\dim W_{s},\dim W_{R})$.
One can then prove that the partial trace of $\left\vert \alpha\right\rangle
\left\langle \beta\right\vert $ (up to the proportionality constant $N$) is an
\textit{unitary} operator $\hat{U}_{\left\langle \beta\right\vert
\otimes\left\vert \alpha\right\rangle }$: (proof in appendix).%

\[
\hat{U}_{\left\langle \beta\right\vert \otimes\left\vert \alpha\right\rangle
}=Ntr_{R}\left(  \left\vert \alpha\right\rangle \left\langle \beta\right\vert
\right)  \text{\qquad with}\qquad\hat{U}_{\left\langle \beta\right\vert
\otimes\left\vert \alpha\right\rangle }^{\dag}=\hat{U}_{\left\langle
\alpha\right\vert \otimes\left\vert \beta\right\rangle }%
\]

Is is tempting to identify somehow the "whole Universe $W=W_{s}\otimes W_{R}$"
in this case with the timeless \textit{unus mundus} of Jung and Pauli.

\section{Quantum Cognition}

The emerging field of quantum cognition advocates the use of a mathematical
formalism inspired by quantum mechanics in order to model certain human
cognitive structures and other complex phenomena. Most researchers of quantum
cognitive sciences are rather pragmatic and careful about the world view
behind their practice \cite{khrennikov2014ubiquitous}, \cite{aerts2009quantum}%
, \cite{busemeyer2012quantum}. They make it clear that they do not assume that
the human brain is quantum-like as claimed by a few scientists
\cite{penrose1994shadows}. Their starting point is the assumption that some
systems (as certain human cognitive and psychological features) have such a
degree of complexity, that it is impossible to study its behavior without
taking into account \textit{contextuality} in the form of non-Kolmogorovian
probability models like the one provided by quantum physics. Many applications
have been pursued successfully for subjects as decision theory and human
judgement in general, conceptual composition, linguistics and memory
recognition. Recently, Dieterik Aerts, after a thorough investigation on how
concepts can compose and interfere with each other in a "quantum-like" manner,
advanced an intriguing proposition that contrasts strikingly from the
prevailing view \cite{aerts2010interpreting}. He set forward the position that
\textit{quantum entities} may actually be\ nothing else than
"\textit{conceptual entities}" themselves. One argument for this is the
striking parallel between the non-local, non-causal \textit{quantum channel}
of two entangled quantum systems and the idea of "\textit{meaning}" of the
composition of two distinct concepts. He further theorizes that space (or
spacetime) is a macroscopic emergent structure and the so-called non-locality
of entangled elementary particles are a consequence of the fact that they
somehow "exist \textit{beyond} space and time". This view resonates obviously
with some of the ideas that we discussed in previous sections. Aerts also
discusses two widely different metaphysical interpretations of this worldview:
The \textit{first} (which he chooses to call an "antropomorphic"
interpretation) views quantum entities and their interactions as an extreme
instance of conceptual entities interacting through "meaningful communication"
embodied by other quantum entities (and their compositions) just as like
humans communicate concepts through ordinary language. A \textit{second
}interpretation that Aerts asserts to be a somewhat less radical and less
antropomorphic philosophical position is a \textit{semiotic} version of these
ideas: here, the quantum entities are nothing but \textit{signs} exchanged
between macroscopic classical systems like the communication between living
creatures and also between inanimate objects as computer interfaces. Again, we
believe that we have here the possibility of an intermediate choice. If one
considers the concept of \textit{consciousness} as a primary ontological
feature of reality then the danger of an antropomorphic interpretation
vanishes. After all, \textit{human} consciousness may be only \textit{one
instance} of the general phenomena of consciousness. Thus, the human species
does not need to play any special role in the nature of reality.

Another interesting paper on quantum cognition is that of Blutner\textit{ et
al} where a quantum-like model is proposed for the structure of Jung's
psychological type\ theory. This is one of the most important parts of Jung's
concept of the \textit{structure of the self \cite{blutner2010two},
\cite{jung1971psychological}, \cite{jung2014two}}. The authors consider a
two-qubit structure where one of the qubits accommodate a two-dimensional
representation of the \textit{introversion/extroversion} attitude. When a
person faces an object from the world, the introvert personality tends to
direct his libido inwards (towards the images that the object elicits in the
subjective world) and the extrovert directs his libido towards the object. The
second qubit models the four psychological functions (\textit{thinking,
feeling}) and (\textit{sensing, intuition}) as two mutually unbiased
orthogonal bases. The \textit{first} pair of opposite functions
\textit{assesses} and \textit{judges} information either \textit{logically} or
\textit{emotionally}. The \textit{second} pair involves \textit{perception} of
information, either \textit{sensorially} or \textit{intuitively}. According to
Jung, every human being is born with a \textit{primary function} which is
independent of race, gender, social class or any other cultural context. A
secondary or auxiliary function necessarily \textit{has} to be sorted out from
the other pair in order to "support" the primary function. Clearly the third
and fourth choices of psychological functions are then uniquely determined.
The third and fourth functions must respectively be the opposites of the
secondary and primary functions. The "upper" hemisphere of this Bloch sphere
is related to the conscious realm while the "lower" hemisphere represents the
unconscious. It follows that the eight possible choices of the psychological
functions together with the two attitude types implies a total of sixteen
different psychological types. The structure of an individual personality type
is therefore described by a state vector belonging to a four-dimensional space
given by the tensor product of these two qubits. Blutner \textit{et al} argue
that the richer topology of the two-qubit model is capable of a much more
adequate and refined explanation of the results obtained from some well-known
methodologies for psychological tests designed to classify which Jungian type
a particular subject belongs to. Jung claims that the psychic dynamics of a
human being takes place through the exchange of libido (psychic energy)
between the conscious and unconscious realms. The primary and secondary
functions belonging to the first realm and the remaining "inferior functions"
related to the latter. A normal and healthy subject must have a balanced and
integrated relation between these distinct structures and the role of the
analyst is to bring back this balance and develop their integration through
the process of individuation. One cannot achieve this directly from the
superior function to the inferior one. It must always be intermediated through
the secondary function. Thus, the path is from the superior to the secondary
function and from there to its opposite. This is the first phase of an
individuation which may be conducted with the help of an analyst. Yet, a full
integration of the superior and inferior functions and therefore the
integration of unconscious elements with the conscious is a personal life-time
task. Jung also speaks of a further classification of the psychological
functions depending on rather they are or not of a fundamentally
\textit{static} or \textit{dynamic} nature. These four "kinds of realities" as
he coined them are the \textit{static reality} that comes through
\textit{sensation}, the \textit{dynamic reality} revealed by
\textit{intuition}, the \textit{static images} provided by \textit{thinking}
and the \textit{dynamic images} perceived by \textit{feeling}. As an example,
consider the static images generated by thinking. These are the timeless
\textit{Logos }of Platonic idealized world of perfect ideas. Yet, feeling has
a superficial resemblance to the thinking process. Let us consider some
illustrative examples chosen by Jung himself. For instance, take the concept
of \textit{freedom}. It can be a highly abstract and static concept but it
also can convey a strong (potentially dynamic) feeling. In a similar manner,
consider the concept \textit{my country}. It clearly can also be taken either
abstractly or emotionally. Another example is the abstract idealized and
static definition of \textit{God} as the\ "unchanging totality of all changing
processes" or rather imagine \textit{God} as a powerful dynamic image
identified with \textit{Eros}. It may prove to be worthwhile to design and
implement some cognitive tests on concept composition like those conducted by
Aerts and others but also accounting somehow for the psychological type of the
subjects. Aerts also observed the complementarity of the opposing ideas of
\textit{abstractness} versus \textit{concreteness} in his analogy between
quantum and conceptual entities. He asserts that the level of "abstractness"
of a given concept is somewhat the measure of its \textit{generality}. For
instance, the concept \textit{animal} is more abstract than the concept
\textit{dog} which in turn is more abstract than \textit{my dog}. The concept
\textit{mine} in this last example was used to give a \textit{context} to the
concept \textit{dog }in the sense that it provides further restrictions for
the examples of \textit{dogs} that can come to a subject's mind. Aerts
compares the opposition of the \textit{abstract/concrete} concept with the
complementarity between position and momentum of quantum particles. Indeed,
the more concrete a concept is, the more "individuality" the concept conveys.
In fact, the more restrictions (contexts) are imposed over it, the more the
level of concreteness of the given concept will increase. Plausibly this could
always be carried on until the ultimate level of concreteness is reached when
"every mind" should agree on its "uniqueness". We have then an individual object.

How can this conception relate to Jung's type theory? It is important to
recognize here that the quantum-like modeling of conceptual space by Aerts and
collaborators entails an impersonal view of the "collection of human minds"
that deals with these concepts. On the other hand, Jung's psychological type
theory seeks the means to understand specific subjects, not only their "minds"
but the complete structure of a subject's\ "self". There also seems to be a
kind of complementarity between these two approaches. Indeed, one approach
studies the space of concepts determined by the "general mind set" of human
beings. The other focuses on the specific ways the psychic structure of an
individual person perceives and accesses concepts and images that are caused
by external objects. From a naive point of view, one may think that the
decision to study both processes within the same mathematical structure would
be a methodological mistake. But the beauty of this conception is the fact
that this allows one to describe within this same quantum-like structure
both\ "the things that are thought" together with "those things that think
them up". This should be expected when one realizes that this division must be
(in a certain sense) an arbitrary splitting of the timeless \textit{unus
mundus} between an observed world and the world that observes it.

Jung firmly believed that causality alone is not sufficient to describe the
full phenomena of the psychology of the self. He asserted that the psychic
structure of the self is a self-regulating system with purposiveness. We
wonder if there is a chance that this teleological characteristic may be
related to the two-state boundary formulation of the \textit{unus mundus}.

\section{Closing Remarks and Future Prospects}

We have tried to put forward some thoughts and speculations on how one could
approach a unified and scientific panpsychist approach to the problem of
consciousness and the emergence of time (and spacetime).

For this goal, we have discussed some attributes from the emerging field of
quantum cognition combined with the world view of analytical psychology
developed by Carl Jung (and to an extent Wolfgang Pauli).

Recent work of Aerts, Blutner and collaborators are fundamental to this
proposed program. It is clear to our mind that the burden of scientific proof
lies on the side of those that advocate the panpsychist view and we hope that
the ideas presented here may somehow help to reach this objective.

A possible future program is the search for a model of the \textit{dynamics}
of the\ "structure of the self" discovered by Jung. The physicist's approach
to model the dynamics of a quantum system is to discover a time evolution rule
(to find a Hamiltonian) either empirically or stemming from some more
fundamental theory. It is our opinion that, in the case of the structure of
the self, it would be constructive to realize that one should rather seek for
a model with the notion of an emerging time (or psychological time) as we have discussed.

Presumably this should be accomplished together with the "individuation" of a
particular subject starting from an undifferentiated unus mundus identified
with the collective unconsciousness. One can also envisage an empirical path
towards this goal. The testing of a psychological sense of time of subjects
under some mind-altered states of consciousness could be interesting.
Consider, for instance, how two persons usually communicate by conversation.
This only occurs because they share a common language and vocabulary.
Communication often becomes easier when these persons know each other and have
developed a sharper set of common words and cultural expressions as jargon,
slangs, etc. It is a familiar feeling that sometimes in a close relationship
one almost can "guess" the others words or thoughts. It is fair to say that in
these cases one could state that there is almost an "extension of the ego" of
each one of the subjects in prejudice to each other's individual ego-centered
conscious. After all, a person's ego is whatever this person identifies itself
with. It can be his body, his job, an ideal, his family, his country, etc. The
ego construct is continuously changing and it can even be determined by the
actual momentarily relationship with another person like talking to a friend,
for instance. Suppose these two friendly subjects are submitted to a test
where they are both initially under the influence of an \textit{extreme}
altered state of conscious. Plausibly at the beginning moments there will not
occur any \textit{conscious} communication between them since each one will be
deeply immersed into their own personal unconscious and even their shared
collective unconscious together with a great decrease of an ordinary sense of
flux of time. Gradually, one expects that some sparse attempts of conscious
communication will happen and the frequency of these ephemeral attempts
probably would slowly increase under the initially mild conscious-will of each
subject together with the withering effects of their mind-altered states. An
impartial noninvasive and thorough observation of this process may provide an
important insight of the inner functioning of some plausible kind of "shared
individuation" of the two subjects. The struggle to communicate and the
gradual increase of common words and concepts may slowly allow the
construction of some kind of "clock ritual" between them. One expects that
this progressively feeling of individualization should be accompanied with a
conjoint perception of a psychological time. Many other ingenious cognitive
experiments may be devised in order to study time perception under altered
states of consciousness and the loss and gain of individuality. Multiple
cognitive time scales of the functioning of the brain have been proposed as an
interplay between quantum-like and classical random electrodynamic signals
from neurons \cite{khrennikov2008quantum}, \cite{khrennikov2011quantum}. Also,
some experiments have already been conducted to study the bistability of
perception under ambiguity \cite{atmanspacher2004quantum}. This maybe a
particularly interesting venue of investigation with regard to Jung's
psychodyamics of mental pathologies because it is well-known that one of the
characteristics of neurotic behavior is the incapability to deal with
ambiguity \cite{weil1986natural}.

\section{Acknowledgments}

Many of the ideas presented in this manuscript had there inception during the
2012-2013 visiting scholarship at the \textit{Institute for Quantum
Studies-Chapman University}. I am particularly grateful to Jeff Tollaksen and
Yakir Aharonov for that great opportunity. I would also like to acknowledge
the \textit{Perimeter Institute} in Canada for the wonderful week in June 2016
at the \textit{Concepts and Paradoxes} conference. I am grateful to Fl\'{a}vio
Cassino, Jonas Cremasco, Pedro Dieguez and Holger Hofmann for many
enlightening discussions on some of the subjects presented here. The author
also acknowledges the support of CNPq.

\section{Appendix}

Given two distinct \textit{maximally entangled} states $\left\vert
\alpha\right\rangle $, $\left\vert \beta\right\rangle $ of $W_{s}\otimes
W_{R}$, we may suppose without loss of generality that $\dim W_{s}=N<\dim
W_{R}=M$ so that we can expand these states as%

\[
\left\vert \alpha\right\rangle =\frac{1}{\sqrt{N}}\sum_{i}\left\vert
u_{i}\right\rangle \otimes\left\vert w_{i}\right\rangle \qquad\text{and\qquad
}\left\vert \beta\right\rangle =\frac{1}{\sqrt{N}}\sum_{j}\left\vert
v_{j}\right\rangle \otimes\left\vert t_{j}\right\rangle
\]
where $\left\{  \left\vert u_{i}\right\rangle ,\left\vert v_{\sigma
}\right\rangle \right\}  $ and $\left\{  \left\vert w_{i}\right\rangle
,\left\vert t_{\sigma}\right\rangle \right\}  $ ($i=0,...,N-1$ and
$\sigma=0,...,M-1$) are pairs of orthonormal basis respectively in $W_{s}$ and
$W_{R}$. We have then that:%

\[
\left\vert \alpha\right\rangle \left\langle \beta\right\vert =\frac{1}{N}%
\sum_{i,j}\left\vert u_{i}\right\rangle \left\langle v^{j}\right\vert
\otimes\left\vert w_{i}\right\rangle \left\langle t^{j}\right\vert
\]
the partial trace gives us then%

\[
\hat{U}_{\left\langle \beta\right\vert \otimes\left\vert \alpha\right\rangle
}=\sum_{i,j}\left\vert u_{i}\right\rangle \left\langle t^{j}\right\vert
w_{i}\rangle\left\langle v^{j}\right\vert \text{\qquad and}\qquad\hat
{U}_{\left\langle \alpha\right\vert \otimes\left\vert \beta\right\rangle
}=\sum_{k,l}\left\vert v_{k}\right\rangle \left\langle w^{l}\right\vert
t_{k}\rangle\left\langle u^{l}\right\vert
\]
then an easy computation proves indeed that%

\[
\hat{U}_{\left\langle \beta\right\vert \otimes\left\vert \alpha\right\rangle
}\hat{U}_{\left\langle \alpha\right\vert \otimes\left\vert \beta\right\rangle
}=\hat{I}\text{\qquad with\qquad}tr\left(  \hat{U}_{\left\langle
\beta\right\vert \otimes\left\vert \alpha\right\rangle }\right)
=N\left\langle \beta|\alpha\right\rangle
\]

\bibliographystyle{unsrt}
\bibliography{acompat,referencias}

\end{document}